# Investigation of Surface Plasmon Resonance in Super-Period Gold Nanoslit Arrays

**Junpeng Guo[*] and Haisheng Leong**

*Department of Electrical and Computer Engineering*

*University of Alabama in Huntsville, Huntsville, AL 35899, USA*

*\*Corresponding author: guoj@uah.edu*

Surface plasmon resonance in super-period metal nanoslits can be observed in the first order diffraction as well in the zeroth order transmission. In this paper, surface plasmon resonance modes in various super-period gold nanoslit arrays are investigated. It is found that the surface plasmon resonance frequencies are determined by the small period of the nanoslits in super-period nanoslits. The number of nanoslits in the unit cell super-period and the nanoslit width do not control the surface plasmon resonance frequencies. It is also found that the resonance wavelength observed in the first order diffraction reveals more accurate the real surface plasmon resonance wavelength in the metal super-period nanoslit array device.

## 1. Introduction

Surface plasmon resonance in metal nanostructures has been a subject of extensive investigations over the past decade. One of most investigated surface plasmon resonance nanostructures is periodic metal nanoslit array [1-12]. A periodic metal nanoslit array supports surface plasmon resonance when it is illuminated with a tranverse magnetic (TM) optical wave that is polarized perpendicularly to the metal nanoslits at the surface plasmon resonance frequency. At the surface plasmon resonance frequncy, a TM optical wave can resonantly tunnel through the nanoslits with significantly enhanced light transmission. Recently, we demonstrated a new surface plasmon resonance spectrometer sensor with a super-period metal nanoslit array without using an optical spectrometer [13]. In a super-period mertal nanoslit array, there are two periods in the nanostructure: one is a small subwavelength period that supports local surface plasmon resonance; another is a large period, called "super-period," that helps to radiate surface plasmon from the metal nanostructure to the far-field. Although a surface plasmon resonance spectrometer sensor based on super-period nanoslits has been experimentally demonstrated, surface plasmon resonances in super-period metal nanoslit structures have not been well invetsigated. In this paper, we present our numerial investigations on various super-period gold nanoslit structure for the purpose to understand how surface plasmon resonances are related to different metal nanoslit structure parameters.

## 2. Structure Dependence of Surface Plasmon Resonances in Super-Period Metal Nanoslit Gratings

A super-period metal nanoslit array consists of periodically arranged regular periodic nanoslit arrays in a thin metal film on a transparent dielectric substrate, as shown in the Fig. 1. The metal nanoslits have a small period p which is below the resonance wavelength of interest. The small period nanoslits are arranged with a large period $P$ by periodically removing one nanoslit for every N nanoslits, where N is an integer number. Once the metal nanoslit device is illuminated, surface plasmons in the metal nanoslits radiate coherently and form non-zeroth order diffractions that propagate in the free space in addition to the zeroth order transmission. At the normal incidence, the angle ($\theta$) of the first order diffraction is $Sin(\theta)=\lambda/P$, where $\lambda$ is the free space wavelength and $P$ is the grating period.

Using the recursive matrix algorithm in the reference [14], we can calculate the zeroth order transmission and the first order diffraction versus the excitation wavelength from the super-period gold nanoslit array shown in the Fig. 1. The small nanoslits period is $p$. The nanoslit aperture width is $w$. The gold film thickness is $t = 60$ nm in all our calculations. The incident optical wave has the polarization perpendicular to the metal nanoslits and is incident normally from the glass substrate. The direction of incidence is designated as the z-direction. The x-direction is perpendicular to the nanoslits. In our calculations, the electrical permittivity of gold film is obtained from a Lorentz-Drude (LD) model based on the experimental result [15]. At the beginning of the calculations, we gradually increase the number of Fourier modes to test the convergence of the calculation results. Finally we include a total number of 170 Fourier modes in our calculations to give accurate results.

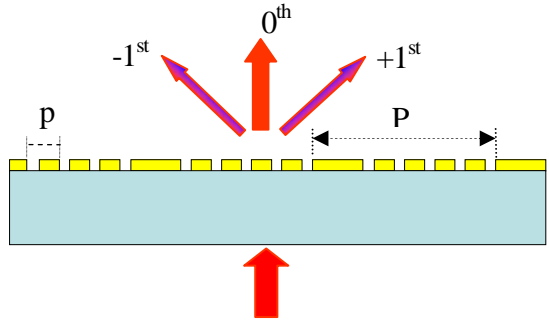

Fig. 1. (Color online) A super-periodic gold nanoslit array with a small period (p) and a large period (P).

First, we keep the nanoslit aperture width of 140 nm to calculate the zeroth order transmittance and the first order diffraction spectra for devices with different small nanoslit periods. The number of nanoslits in each super-period is four, i.e. every five small periods, one nanoslit is missing. The gold film thickness is 60 nm. The spectral resolution in our calculations is 1.0 nm. Fig. 2 (a) shows the zeroth order transmission spectra for devices of three different nanoslit periods of 390 nm, 420 nm, and 450 nm, respectively. Fig. 2 (b) shows the first order diffraction efficiencies from the three devices. For the super-period nanoslit device with the small period of 390 nm, the zeroth order transmission peak is at the 575 nm wavelength as shown in the Fig. 2(a). The corresponding first order diffraction peak wavelength is at 566 nm which is close but slightly smaller than the zeroth order tranmission peak wavelength. The enhanced zeroth order transmission and the corresponding first order diffraction peak are due to the surface plasmon resonance in the 390 nm period metal nanoslits. For the super-period nanoslit device with 420 nm small period, the maximal zeroth order transmission is at the 612 nm wavelength. The corresponding first order diffraction peak is at 609 nm which is close but slightly less than the zeroth order tranmission peak wavelength. The enhanced zeroth order transmission and the corresponding first order diffraction peak are due to the surface plasmon resonance in the 420 nm period metal nanoslits. For the super-period nanoslit device with the 450 nm small period, the maximal zeroth order transmission is at the 653 nm wavelength. The corresponding first order diffraction peak is also at 653 nm. The maximal zeroth order transmission and the corresponding first order diffraction peak at this wavelength are due to the surface plasmon resonance in the metal nanoslits. It can be seen that the surface plasmon resonance wavelength increases as the small period increases and the resonance wavelength observed in the first order diffraction is about the same, but not exactly the same as the resonance wavelength observed in the zeroth order transmission.

To compare with the surface plasmon resonance in regular metal nanoslit arrays, we calculated the transverse magnetic (TM) optical wave transmittance through regular gold nanoslit arays with period of 390 nm, 420 nm, and 450 nm on the same kind of substrates. The gold film thickness is 60 nm. The calculation results are plotted in Fig. 2(c), which show enhanced transmissions at the surface plasmon resonance wavelengths. The peak transmission wavelength is the same as the peak wavelength in the first order diffraction from the super-period nanoslit array. Therfore, the resonance wavelength observed in the first order diffraction from the super-period metal nanoslit array reveals better the surface plasmon resonance in the regular metal nanoslit array.

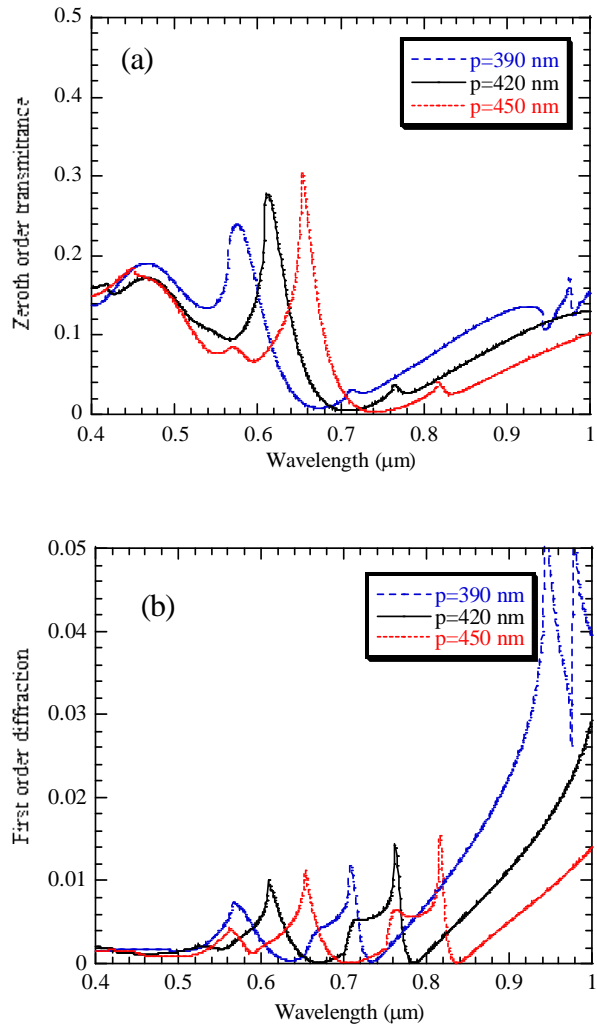

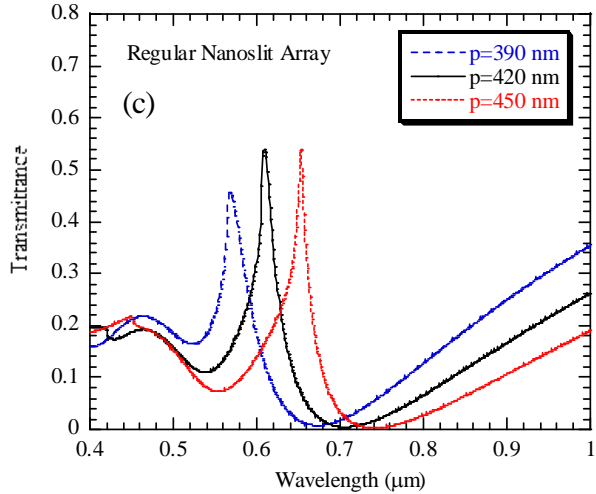

Fig. 2. (Color online) (a) The zero-order transmission spectra and (b) the first order diffraction spectra from the super-period nanoslit arrays of three different nanoslit small periods and the same slit width (w=140 nm)., (c) transmittance through regular periodic gold nanoslit arrays with the different small periods of 390 nm, 420 nm, and 450 nm.

Super-period metal nanoslit arrays have additional surface plasmon resonance modes comparing with the regular metal nanoslit arrays. The additional surface plasmon resonance modes are due to the large period in the grating structure. For the super-period metal nanoslit array with 420 nm small period and 2100 nm super-period, the fundamental surface plasmon resonance is at 609 nm. Another surface plasmon resonance is at 672 nm. To understand the surface plasmon resonance in the super-period metal nanoslits, we calculated the electric field distributions at the wavelength of 609 nm. Fig. 3(a) shows the Ex component of the electric field distribution in the device at the 609 nm wavelength in the log scale. Fig. 3(b) shows the Ez component of electric field distribution in the device at the 609 nm wavelength in the log scale. Fig. 3(c) shows the Ex component of the electric field distribution in the device at the peak zeroth order transmission wavelength of 612 nm in the log scale. Fig. 3(d) shows the Ez component of electric field distribution in the device at the peak zeroth order transmission wavelength of 612 nm in the log scale. It can be seen clearly that the electric field in the device is stronger at the first order diffraction peak wavelength than the electric field at the peak zeroth order transmittance wavelength. This indicates the the resonance observed in the first order diffraction reveals more accurate the surface plasmon resonance occured in the super-period metal nanoslit array device.

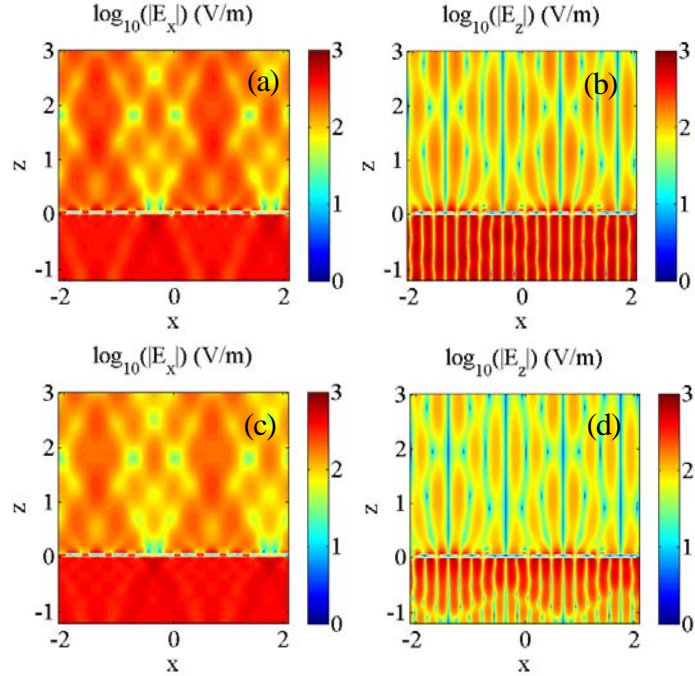

Fig. 3. (Color online) The electric field distributions in the super-period metal nanoslits array in the log scale: (a) Ex component of the electric field at the peak first order diffraction wavelength of 609 nm; (b) Ez component of the electric field at the peak first order diffraction wavelength of 609 nm; (c) Ex component of the electric field at the zeroth order peak transmittance wavelength of 612 nm; (d) Ez component of the electric field at the 612 nm wavelength.

Each nanoslit in device can be thought as an optical antenna which radiates upon the optical excitation. The radiation of the nanoslit antenna array reaches to the maximum when the surface plasmon resonance is excited. The nanoslit antennas array gives a radiation pattern as it is seen the in the spatial distribution of the Ex component. The electric field is focused at a distance of approximate 1.2 micron above the nanoslit metal surface. This is due to the interference of radiations from the four nanoslits in each sub-array. From the Fig. 3 (b) and Fig. 3(c), it can be seen that the Ez component is zero in the center lines of 700 nm wide metal strips and also in the center lines of each nanoslit unit cells. This is due to the symmetry of the super-period nanoslit structure and the Ez becomes zero along these lines. Fundamentally, it is the interference of nanoslit radiations that give the field distribution pattern. The surface plasmon resonance in the nanoslit array enhances the radiation.

Fig. 4 (a) and (b) show the electric field distributions of the Ex component and the Ez component respectively at 762 nm, the first order diffraction peak wavelength. The metal nanoslits on both sides of the 700 nm wide metal strips radiate more strongly than the nanoslits in the center of the nanoslits array. This indicates that the surface plasmon resonance in 700 nm wide metal strips enhances the radiation of the nanoslits dipole antennas on both sides of the gold metal strips. In Fig. 4(a), it can be seen that the Ex field is focused in a distance of about 1.5 micron above the nanoslits metal surface. The further focal distance than that in the Fig. 3(a) is due to the longer wavelength of the surface plasmon reasonance. Fig. 4(c) shows the Ex component of the electric field distribution in the device at the peak zeroth order transmission wavelength of 674 nm in the log scale. Fig. 3(d) shows the Ez component of electric field distribution in the device at the peak zeroth order transmission wavelength of 674 nm in the log scale. It can be seen that the electric field in the device is stronger at the first order diffraction peak wavelength than the electric field at the peak zeroth order transmittance wavelength. This indicates the the resonance observed in the first order diffraction reveals more accurate the surface plasmon resonance in the metal nanostructure.

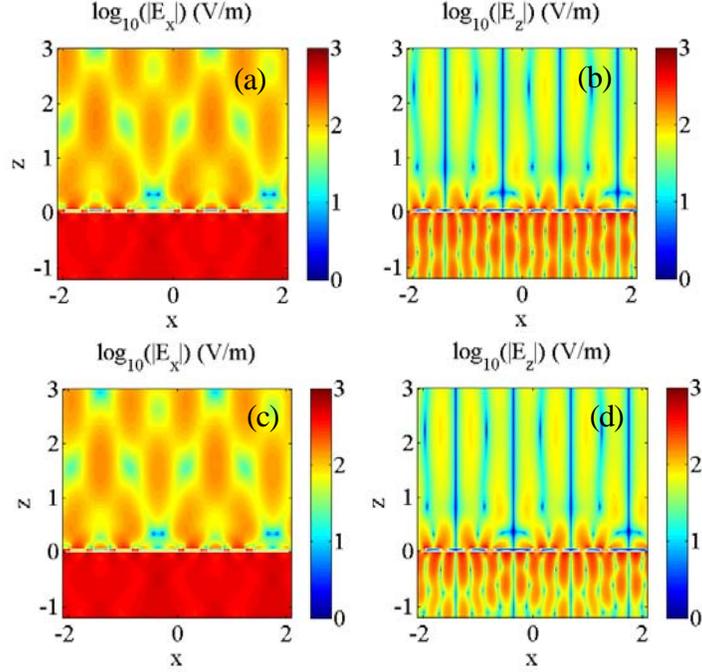

Fig. 4. (Color online) The electric field distributions in the super-period metal nanoslits array in the log scale: (a) Ex component of the electric field at the wavelength of 762 nm, (b) Ez component of the electric field at the wavelength of 762 nm, (c) Ex component of the electric field at the peak zeroth order transmittance wavelength of 764 nm; (d) Ez component of the electric field at the peak zeroth order transmittance wavelength of 764 nm.

We also calculated the zeroth order transmission and the first order diffraction spectra from super-period metal nanoslit arrays of different nanoslits widths ($w$) but the same small nanoslit period of 420 nm and the same large period of 2100 nm. The metal film thickness is at 60 nm. Fig. 5(a) shows the zeroth order transmission spectra for three super-period nanoslit arrays with nanoslit width of 120 nm, 140 nm, and 160 nm, respectively. Fig. 5(b) shows calculated first order diffraction spectra from these three super-period nanoslit arrays. As it is seen in the Fig. 5, the change of nanoslit width has no impact on the resonance wavelength, but has a significant impact on the line-width of the resonance. Narrower aperture width gives narrower line-width and slightly reduced zeroth order transmission and first order diffraction.

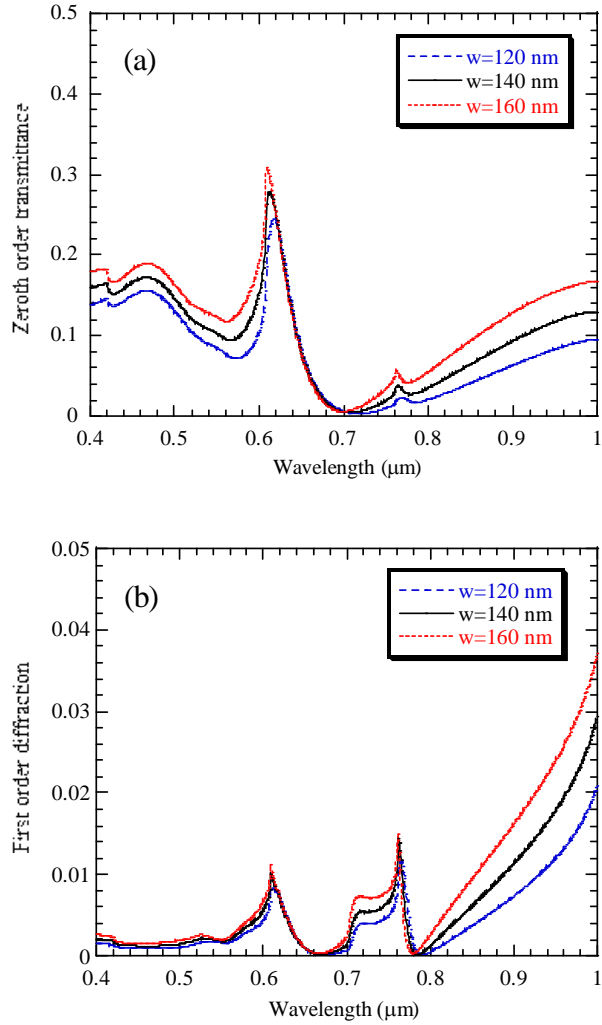

Fig. 5. (Color online) (a) The zeroth order transmittance and (b) the first order diffraction from the super-period gold nanoslits arrays with different nanoslits widths.

To find out how the number of nanoslits in each super-period affects the resonance, we calculated the zeroth order transmission and the first order diffraction for super-period nanoslit arrays with different numbers of nanoslits in the super-period unit cell while keeping all the other parameters same. Fig. 6 (a) shows the zeroth order transmission spectra from the super-period nanoslits with increasing number of nanoslits in the super-period unit cell from three to five. The nanoslit width is 140 nm and the small period is 420 nm for all three devices. The super-period is 1680 nm, 2100 nm, and 2520 nm, respectively. The zeroth order tranmission peaks are all at 612 nm although they have different resonance strengths. The first order diffraction peaks are at 609 nm. It is clearly seen that the resonance wavelength does not change when the number of nanoslits in the super-period unit cell changes. The resonance strength in the first order diffraction decreases as the number of nanoslits per the large period increases as shown in the Fig. 6 (b). This is because less energy resides in the first order diffraction as more diffraction orders start to get in as the super-period increases. We also calculated the zeroth order transmission and the first order diffraction from super-period nanoslit arrays with two nanoslits in the super-period unit cell. Surface plasmon resonances can also be observed in the the zeroth order transmission and in the first order diffraction for the super-period unit cell of two nanoslits.

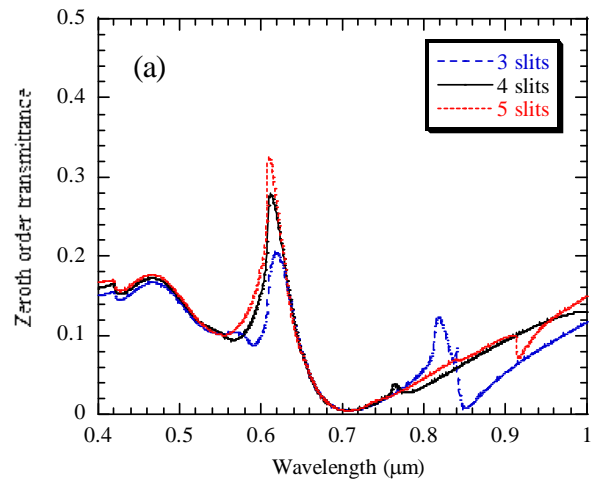

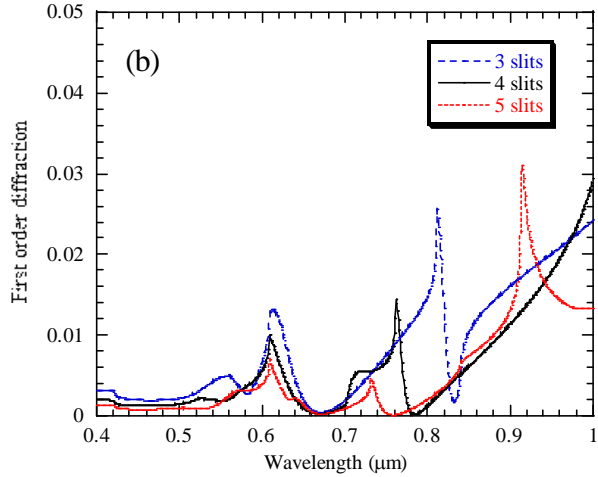

Fig. 6. (Color online) (a) The zeroth order transmission and (b) the first order diffraction of super-period metal nanoslit arrays with different number of nanoslits in the super-period unit cell.

## 3. Summary

In summary, we have calculated the zeroth order transmission and the first order diffraction versus the wavelength from various super-period gold nanoslit arrays upon TM optical wave excitations. It is found that the peak diffraction wavelength in the first order diffraction is very close to but not exactly the same as the zeroth order transmission peak wavelength. The peak wavelength in first order diffraction is the same as the peak transmission wavelength of the regular nanoslit array with the same period. The resonance observed in the first order diffraction spectrum reveals more accurate the surface plasmon resonanc occured in the super-period nanoslit device. It is also found that the change of the number of nanoslits in the unit cell of the super-period nanoslit array does not affect the surface plasmon resonance frequncy. The surface plasmon resonance frequency is primarily controlled by the small period of nanoslits in the unit cell and does not change with the nanoslits width and the number of nanoslits in the supere-period unit cell. The findings in this work provide new insights to the surface plasmon resonance phenomena in super-periog metal nanosslit arrays. Super-period metal nanoslit device invetsigated here can be integrated with nano and microfluidic devices and potentially usefull for many biochemical and biomedical measurement applications.

## Acknowledgment

This work was partially sponsored by the National Science Foundation (NSF) through the award No. NSF-0814103.

## References

1. J. A. Porto, F. J. García-Vidal, and J. B. Pendry, "Transmission resonances on metallic gratings with very narrow slits," Phys. Rev. Lett. 83, 2845-2848 (1999).

2. P. Lalanne, J. P. Hugonin, S. Astilean, M. Palamaru, and K. D. Möller, "One-mode model and Airy-like formulae for one-dimensional metallic gratings," J. Opt. A: Pure Appl. Opt. 2, 48-51 (2000).

3. M. M. J. Treacy, "Dynamical diffraction explanation of the anomalous transmission of light through metallic gratings," Phys. Rev. B 66, 195105.1 – 195105.11 (2002).

4. Z. Sun, Y. S. Jung, and H. K. Kim, "Role of surface plasmons in the optical interaction in metallic gratings with narrow slits," Appl. Phys. Lett. 83, 3021-3023 (2003).

5. M. Dykhne, A. K. Sarychev, and V. M. Shalaev, "Resonant transmittance through metal films with fabricated and light-induced modulation," Phys. Rev. B 67, 195402.1-195402.13 (2003).

6. S. A. Darmanyan, M. Nevière, A. V. Zayats, "Analytical theory of optical transmission through periodically structured metal films via tunnel-coupled surface polariton modes," Phys. Rev. B 70, 075103.1-075103.9 (2004).

7. K. G. Lee and Q. H. Park, "Coupling of surface plasmon polaritons and light in metallic nanoslits," Phys. Rev. Lett. 95, 103902.1-103902.4 (2005).

8. P. Lalanne, J. P. Hugonin, and J. C. Rodier, "Theory of surface plasmon generation at nanoslit apertures," Phys. Rev. Lett. 95, 263902.1-263902.4 (2005).

9. J. Wuenschell and H. K. Kim, "Surface plasmon dynamics in an isolated metallic nanoslit," Opt. Exp. 14, 10000-10013 (2006).

10. D. Pacifici, H. J. Lezec, H. A. Atwater, and J. Weiner, "Quantitative determination of optical transmission through subwavelength slit gratings in Ag films: Role of surface wave interference and local coupling between adjacent slits," Phys. Rev. B 77, 115411.1-115411.5 (2008).

11. Y. S. Jung, J. Wuenschell, H. K. Kim, P. Kaur, and D. H. Waldeck, "Blue-shift of surface plasmon resonance in a metal nanoslit array structure," Opt. Exp. 17, 16081-16091 (2009).

12. G. D'Aguanno, N. Mattiucci, M. J. Bloemer, D. de Ceglia, M. A. Vincenti, and A. Alù, "Transmission resonances in plasmonic metallic gratings," J. Opt. Soc. Am. B 28, 253-264 (2011).

13. H. Leong and J. Guo, "Surface plasmon resonance in super-periodic metal nanoslits," Opt. Lett. 36, 4764-4766 (2011).

14. L. Li, "Formulation and comparison of two recursive matrix algorithms for modeling layered diffraction gratings," J. Opt. Soc. Am. A 13, 1024-1035 (1996).

15. A. D. Rakic, A. B. Djurišic, J. M. Elazar, and M. L. Majewski, "Optical properties of metallic films for vertical-cavity optoelectronic devices," Appl. Opt. 37, 5271-5283 (1998).